\documentclass[aps,prb,twocolumn,showpacs,superscriptaddress,floatfix,nofootinbib]{revtex4}
\usepackage{amsfonts,hyperref}

\usepackage{graphicx}
\usepackage{amsmath}
\usepackage{times}
\usepackage{amssymb}
\usepackage{subfigure}
\usepackage{array}
\usepackage{tikz}
\usetikzlibrary{calc}
\usepackage{float}

\usetikzlibrary{backgrounds}
\tikzset{spin/.style={circle=2pt,draw=black!100,fill=orange!80,inner sep=3pt}}

\usepackage{hyperref}
\hypersetup{
  colorlinks=true,
  pdfauthor={Laura Messio}
}

\begin{document}

\title{Entropy dependence of correlations in one-dimensional SU(N) antiferromagnets}

\author{Laura Messio}
\affiliation{Institut de Physique Th\'eorique (IPhT), CEA, CNRS, URA 2306, F-91191 Gif-sur-Yvette, France}
\author{Fr\'ed\'eric Mila}
\affiliation{Institute of Theoretical Physics, \'Ecole Polytechnique F\'ed\'erale de Lausanne (EPFL), CH-1015 Lausanne, Switzerland}

\date{\today}

\begin{abstract}
Motivated by the possibility to load multi-color fermionic atoms in optical lattices,
we study the entropy dependence of the properties of the one-dimensional antiferromagnetic
$SU(N)$ Heisenberg model, the effective model of the $SU(N)$ Hubbard model with one particle
per site (filling $1/N$). Using continuous-time world line
Monte Carlo simulations for $N=2$ to $5$,
we show that characteristic short-range correlations develop at low temperature as a precursor
of the ground state algebraic correlations. We also calculate the entropy as a function of
temperature, and we show that the first sign of short-range order appears at an entropy per particle
that increases with $N$ and already reaches $0.8k_B$ at $N=4$, in the range of experimentally
accessible values.
\end{abstract}

\maketitle

Lattice $SU(N)$ models play an ever increasing role in the investigation of strongly
correlated systems, both in condensed matter and in cold atoms. The first systematic
use of these models took place in the context of the large-$N$ generalization of the
$SU(2)$ Heisenberg model, in which conjugate (or self-conjugate) representations
are put on the two sublattices of the square lattice so that a $SU(N)$ singlet can be formed on two sites\cite{Affleck_largeN,Sachdev_fermions,Auerbach_largeN}. Over the years,
another class of $SU(N)$ models with the same representation at each site has
 appeared as the
relevant description of the low temperature properties in several contexts. In particular,
the $SU(3)$ model corresponds to the spin-1 Heisenberg model with equal bilinear and biquadratic interactions\cite{laeuchli_2006,toth_2010,toth_2012},
while the $SU(4)$ model is equivalent to the symmetric version of the Kugel-Khomskii
model of Mott insulators with orbital degeneracy\cite{kugel1982,li1998}. These models have however attracted
renewed attention recently as the appropriate low energy theory of ultracold gases of
alkaline-earth-metal atoms in optical lattices in the Mott insulating phase with one atom per site,
the parameter $N$ corresponding to the
number of internal degrees of freedom of the atoms\cite{Nature_SUN}.

A peculiar characteristic of these $SU(N)$ models is that one needs $N$ sites to form a singlet.
This is often reflected in their ground state properties. In one dimension, the $SU(N)$ model has been
solved with Bethe ansatz for arbitrary $N$\cite{Sutherland_SUN}, and the dispersion of the elementary fractional excitations
has a periodicity $2\pi/N$. On a ladder, the $SU(4)$ model has a plaquette ground state\cite{vdb_2004}.
In two-dimensions, the $SU(3)$ model on both the square and triangular lattices has long-range
color order with 3-site periodicity along the lines\cite{laeuchli_2006,toth_2010}, while on the kagome lattice it is spontaneaously
trimerized\cite{corboz_kagome}. The $SU(4)$ model on the checkerboard lattice also has a plaquette ground state\cite{corboz_kagome}.
Even on the square lattice, where the $SU(4)$ model undergoes spontaneous dimerization\cite{corboz_2011} with possibly
algebraic correlations\cite{vishwanath_2009}, neighboring
dimers involved pairs of different colors, so that the 4 colors are indeed present with equal weight
on all plaquettes. The general properties for arbitrary $N$ are not known however. An adaptation of
the previous large-$N$ studies has been proposed for $m$ atoms per site\cite{Hermele_largeN_SUN}. If $m=O(N)$, the
ground state has been proposed to be a chiral spin liquid for large $N$. 

The wealth of ground states predicted for different $N$ on various lattices calls
for an experimental investigation. Ultra-cold fermionic atoms can a priori lead to very accurate
realizations of these models. However, the temperature is a limiting factor.
It can be lowered with respect to the initial temperature if the optical
lattice is adiabatically switched on\cite{entropy_cold_atoms}, but it cannot be made arbitrarily small. In fact, with
adiabatic switching, one can control the entropy rather than the temperature, and in current
state-of-the-art experimental setups, the lower limit for fermions with $N=2$ is equal to $0.77 k_B$ per particle\cite{jordens_2010}. If contact
is to be made with experiments on cold atoms, it is thus crucial to know the properties
of a given model as a function of entropy. For the SU(2) Heisenberg model on the cubic lattice, N\'eel ordering takes place at an entropy 0.338 $k_B$, i.e. about half the value that can be achieved today\cite{jordens_2010}.

The first hint that increasing the number of colors might help in beating this experimental limit has been obtained in the context of a high temperature investigation of the $N$-flavour Hubbard model by Hazzard et al\cite{hazzard_2012}, who have shown that the effective temperature reached after introducing the optical lattice decreases with $N$ under fairly general conditions. However, to the best of our knowledge, no attempt has been made so far to determine how the temperature or the entropy below which signatures of the ordering will show up depends on $N$.

In this Letter, we address this issue in the context of the one-dimensional (1D) antiferromagnetic $SU(N)$ Heisenberg model on the basis of extensive Quantum Monte Carlo (QMC) simulations.
As we shall see, the ground state algebraic correlations lead to characteristic anomalies in the structure factor upon lowering the temperature. These
anomalies only become visible at quite low temperature, but remarkably enough, the corresponding entropy per
particle increases with $N$, leading to observable qualitative effects with current experimental setups for $N\ge 4$.

{\it The $SU(N)$ Heisenberg model.---} A good starting point to discuss $N$-color fermionic atoms loaded in an optical lattice is the $SU(N)$ Hubbard model defined by the Hamiltonian:
  \begin{equation}
   \widehat H=t\sum_{\langle i,j\rangle\alpha}(\widehat c^\dag_{\alpha i}\widehat c_{\alpha j}+h.c.)+U\sum_{i,\alpha<\beta}\widehat n_{\alpha i}\widehat n_{\beta i},
   \label{eq:Ham_Hubbard}
  \end{equation}
  where $\widehat c^\dag_{i,\alpha}$ and $\widehat c_{i,\alpha}$ are creation and annihilation operators of a fermion of color $\alpha=1\dots N$ on site $i$ and the sum is over the first-neighbors of a periodic chain of length $L$.
  $\widehat n_{\alpha i}$ is the number of fermions of color $\alpha$ on site $i$. At filling $1/N$, i.e. with one
  fermion per site, the ground state is a Mott insulator, and to second order in $t/U$, the low-energy effective Hamiltonian is the $SU(N)$ Heisenberg model with the fundamental $SU(N)$ representation at each site, and with
  coupling constant $J=2t^2/U$. Setting the energy unit by $J=1$, this Hamiltonian can be written (up to an additive constant):
  \begin{equation}
   \widehat H= \sum_{\langle ij\rangle} \widehat P_{ij}.
   \label{eq:Ham}
  \end{equation}
  where $\widehat P_{ij}$ permutes the colors on sites $i$ and $j$. If we denote by
  $\widehat S^{\alpha\beta}_i$ the operator that replaces color $\beta$ by $\alpha$ on site $i$, this permutation
  operator can be written as:
  \begin{equation}
   \widehat P_{ij}=\sum_{\alpha,\beta} \widehat S^{\alpha\beta}_i\widehat S^{\beta\alpha}_j
   \label{eq:Pij}
  \end{equation}

This effective Hamiltonian is an accurate description of the system provided the temperature is much smaller than
the Mott gap. In terms of entropy, the criterion is actually quite simple. The high temperature limit of the entropy per site of the $SU(N)$ Hubbard model at $1/N$-filling can be shown to be equal to $k_B (N \ln N - (N-1) \ln (N-1))$, while that of the $SU(N)$ Heisenberg model is equal to $k_B \ln N$. So we expect the description in terms of the Heisenberg model to be accurate when the entropy is below $k_B \ln N$. For $SU(2)$, this is a severe restriction for experiments since $\ln 2\simeq 0.693...$, but already for $SU(3)$, this is less of a problem since $\ln 3 \simeq 1.099$. Of course, this
is not the whole story since what really matters is the entropy below which specific correlations develop, but this
is an additional motivation to consider $SU(N)$ models with $N>2$.

{\it Exact results.---} A number of exact results that have been obtained over the years
on the 1D $SU(N)$ Heisenberg model will prove to be useful.
The model has been solved with Bethe ansatz by Sutherland\cite{Sutherland_SUN}.
He showed that, in the thermodynamic limit, the energy per site is given by
  \begin{equation}
   E_0(N)=2\sum_{k=2}^\infty\frac{(-1)^k\zeta(k)}{N^k}-1,
  \end{equation}
  where $\zeta$ is the Riemann's zeta function. Some values are given in Tab.~\ref{tab:energies_SUN}.
  In addition, he showed that there are $N-1$ branches of elementary excitations which all have the same velocity
  $v=2\pi/N$ at small $k$. Affleck has argued that the central charge $c$ should be equal to $N-1$\cite{affleck_1988}, and Lee has shown
  that\cite{Lee_SUN}, at low temperature $T$, the entropy is given by:
\begin{equation}
    S(T)=\frac{k_BN(N-1)}{6}T+O(T^2),
    \label{eq:entropy_slope}
  \end{equation}
  a direct consequence of $c=N-1$ and $v=2\pi/N$ since the linear coefficient is equal to $\pi c/3v$.


{\it The QMC algorithm.---} Quantum Monte-Carlo is the most efficient method to study the finite temperature
properties of interacting systems provided one can find a basis where there is no minus sign problem, i.e.
a basis in which all off-diagonal matrix elements of the Hamiltonian are non-positive. For the $SU(2)$ antiferromagnetic Heisenberg model on bipartite lattices, this is easily achieved by a spin-rotation by
$\pi$ on one sublattice. For $SU(N)$ with $N>2$, there is no such general solution, but in 1D one can get
rid of the minus sign on a chain with open boundary conditions, as already noticed for the SU(4) model\cite{frischmuth_1999}. Let us start from the natural basis consisting of the $N^{L}$ product states $\otimes_i|\alpha_i\rangle=|\alpha_0,\dots,\alpha_{L-1}\rangle$, where $\alpha_i$ is the color at site $i$.
In this basis, all off-diagonal elements of the $SU(N)$ model of Eq.\ref{eq:Ham} are either zero or positive.
However, a generalization of the Jordan-Wigner transformation allows to change all these signs on an open chain.
This transformation is defined by:
\begin{equation}
   |\alpha_0,\dots,\alpha_{L-1}\rangle \to (-1)^{r(\alpha_0,\dots,\alpha_{L-1})}|\alpha_0,\dots,\alpha_{L-1}\rangle,
  \end{equation}
  where $r(\alpha_0,\dots,\alpha_{L-1})$ is the number of permutations between different color particules on neighboring sites needed to obtain a state such that the $\alpha_i$ are ordered ($\alpha_i\leq\alpha_j$ for $i<j$).
  This basis change is equivalent to a Hamiltonian transformation, the new Hamiltonian being given by:
  \begin{equation}
   \widehat H=\sum_{\langle ij\rangle}\sum_{\alpha} \left( \widehat S^{\alpha\alpha}_i\widehat S^{\alpha\alpha}_j- \sum_{\beta\neq\alpha} \widehat S^{\alpha\beta}_i\widehat S^{\beta\alpha}_j\right).
   \label{eq:Ham2}
  \end{equation}
On a periodic chain, the equivalence with the Hamiltonian of Eq.~\ref{eq:Ham} is not exact, but the difference disappears in the thermodynamic limit. So in the following we will simulate the Hamiltonian of Eq.\ref{eq:Ham2}.

To do so, we have developed a continuous time world-line algorithm with cluster updates\cite{QMC} adapted to the model of Eq.~\ref{eq:Ham2} with $N$ colors.
The partition function $Z$ is expressed as a path integral over the configurations $\phi:\tau\to\phi(\tau)$, where $\tau$ is the imaginary time going from 0 to $\beta=\frac1{k_BT}$ and $\phi(\tau)$ is a basis state.
The functions $\phi$ that contribute to the integral can be represented by $\phi(0)$ and by a set of world line crossings $\{(i,j,\tau)\}$ that exchange the colors of two sites $i$ and $j$ at time $\tau$.
  A local configuration $c$ on a link $ij$ at time $\tau$ is represented by
  \begin{equation}
   c=\left(\begin{matrix}
     \alpha_i(\tau^+)\,\alpha_j(\tau^+) \\  \alpha_i(\tau^-)\,\alpha_j(\tau^-)
     \end{matrix}\right)
  \end{equation}
Cluster algorithms are well documented for 2-color models.
Here we generalize the approach to $SU(N)$ by choosing randomly two different colors $p$ and $q$ out of $N$ and by constructing clusters on which only these two colors are encountered. The steps to construct the clusters are the following.
  We first randomly place elementary graphs in the configuration using a Poisson distribution. These graphs are drawn in the first column of Tab.\ref{tab:cluster} and the Poisson time constant is given in the last column. They are accepted only if $\Delta_G(c)=1$ (if a color which is neither $p$ nor $q$ appears in the local configuration, the graph is rejected).
  Then we assign graphs to the world-line crossings between $p$ and $q$ colors using the last two columns of Tab.\ref{tab:cluster}.
  At the places where no graph has been attributed, we follow the path with the same color.
  Finally we follow each constructed cluster and exchange $p$ and $q$ on it with a probablity $1/2$ (Swendsen-Wang algorithm).
  This constitutes a Monte Carlo step.

  \begin{table}
    \begin{displaymath}
      \begin{array}{!{\vrule width 1pt}c!{\vrule width 1pt}c|c|c!{\vrule width 1pt}c!{\vrule width 1pt}}
        \noalign{\hrule height 1pt}
        G& \Delta_G\left(\begin{matrix}
        p\, p \\     p\, p
        \end{matrix}\right)&
        \Delta_G\left(\begin{matrix}
        p\,q \\     p\,q
        \end{matrix}\right)&
        \Delta_G\left(\begin{matrix}
        q\,p \\     p\,q
        \end{matrix}\right)&
        W_G\\
        \noalign{\hrule height 1pt}
        \includegraphics[width=.025\textwidth]{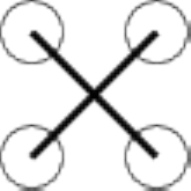}  &1 &- &1 &\epsilon\\
        \includegraphics[width=.025\textwidth]{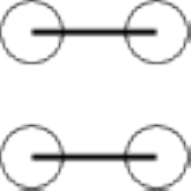}  &- &1 &1 &1-\epsilon\\
        \includegraphics[width=.025\textwidth]{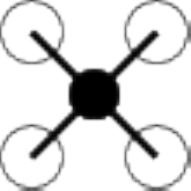}  &0 &1 &0 &2\epsilon\\
        \noalign{\hrule height 1pt}
      \end{array}
    \end{displaymath}
    \caption{
      Fabrication rules for clusters.
      In the first column are drawn all possible graphs $G$.
      The three following columns are $\Delta_G(c)$ equal to 1 when $G$ is compatible with the local configuration $c$.
      The last column is the constant of time used to sample the graphs with a Poisson law.
      $\epsilon$ is a small number ensuring the ergodicity ($\epsilon$ is taken $0.01$ in our simulations).
      \label{tab:cluster}
    }
  \end{table}

Using this algorithm, we have calculated the energy per site $E$, which is given by:
  \begin{equation}
   E
   =\left\langle \frac{\widehat H}{L}\right\rangle
   \simeq\frac{k_BT}{L\,n}\sum_\phi\left(\sum_{\langle ij\rangle}\int d\tau
   \delta_{\alpha_i(\tau),\alpha_j(\tau)}-n(\phi)\right),
  \end{equation}
  where $n$ is the number of Monte Carlo steps and $n(\phi)$ the number of world-line crossings in the configuration $\phi$, the diagonal correlations defined by
  \begin{equation}
  C(j)
  =\left\langle \sum_\alpha \widehat S^{\alpha\alpha}_0 \widehat S^{\alpha\alpha}_j\right\rangle -\frac1{N}
  \end{equation}
and the associated structure factor defined by
  \begin{equation}
  \tilde C(k)=\frac1{2\pi} \frac{N}{N-1}\sum_jC(j)e^{ikj}.
  \end{equation}
This structure factor is normalized in such a way that $\frac{2\pi}{L}\sum_k  \tilde C(k)=1$.

  \begin{figure}
    \begin{center}
      \includegraphics[width=0.48\textwidth]{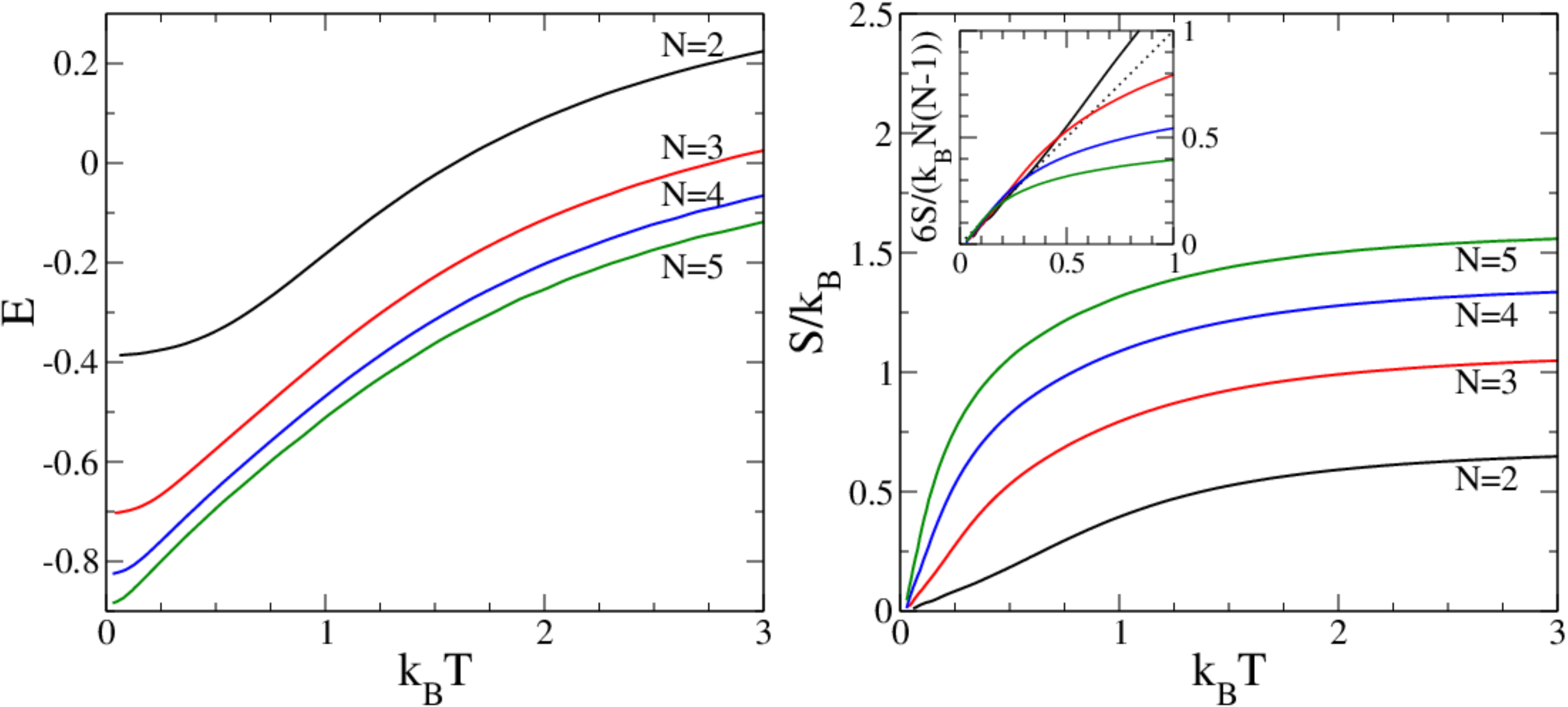}
    \end{center}
    \caption{
    Evolution of the energy per site $E$ and of the entropy per site $S$ as a function of the temperature $T$ for different $N$ on a $L=60$ chain.
    The inset shows the slope of the entropy at $T=0$, given in Eq.~\ref{eq:entropy_slope}.
    The curvature being positive at $T=0$, the curves go higher than the tangent (dashed line).
    \label{fig:fdeTS_chain}}
  \end{figure}

{\it The results.---} We have studied chains of length $L=60$ for $T$ from $0.01$ to $20$ with a number of colors $N=2,3,4$ and $5$, and a number of Monte Carlo steps $n$ at least equal to $10^6$.
  The correlation time measured by the binning method indicates that around $N$ steps are needed to obtain uncorrelated configurations, whatever the temperature, and that
  the precision on the energy per site $E$ is better than $10^{-4}$. This could be confirmed by the comparison of the limit of the energy when $T\to0$ with the exact finite $L$ value for $SU(3)$.\cite{Martins_SUN}
  Moreover, the energy of the ground state differs from that of the thermodynamic limit by less than $8.10^{-4}$.
  So, for our purpose, the finite size effects can be considered to be negligible (see Tab.~\ref{tab:energies_SUN}).
  The entropy per site $S$ has been deduced from the energy $E$ by an integration from high temperature:
  \begin{equation}
   S(T)=S(\infty)-\int_T^\infty d\tau \frac{k_B}{\tau}\frac{dE}{d\tau}
  \end{equation}
  where $S(\infty)=k_B\ln(N)$.
  $E$ and $S$ are plotted in Fig.~\ref{fig:fdeTS_chain} for different $N$ as a function of $T$.
  Since the entropy is the result of a numerical integration, it is important to check its accuracy,
  especially at low temperature since by construction it has to be correct at high temperature.
  Now, we know that, at low temperature, the entropy must be linear with a slope equal to $k_BN(N-1)/6$
  (see Eq.\ref{eq:entropy_slope}). This is confirmed by the inset of Fig. 1b), in which one clearly sees
   that the entropies times $6/k_BN(N-1)$ lie on top of eachother at low temperature.

  Now, the stabilization of the energy at low $T$ occurs at a temperature that decreases when $N$ increases.
  Thus, one could naively think that it will be more difficult to observe the development of the ground state correlations when $N$ increases. However, this is not true if one considers the entropy. Indeed,
  the entropy grows much faster at low temperature when increasing $N$. So, the temperature
  corresponding to a given entropy decreases very fast when $N$ increases.

  \begin{table}
    \begin{displaymath}
      \begin{array}{!{\vrule width 1pt}c!{\vrule width 1pt}c|c|c!{\vrule width 1pt}}
        \noalign{\hrule height 1pt}
        N & BA(L=\infty) & BA(L=60) & QMC(L=60) \\
        \noalign{\hrule height 1pt}
        2& -0.386294  &             &-0.38675(2)\\
        3& -0.703212  &-0.7038228   &-0.70384(2)\\
        4& -0.8251193 &             &-0.82577(2)\\
        5& -0.884730  &             &-0.88541(2)\\
        \noalign{\hrule height 1pt}
      \end{array}
    \end{displaymath}
    \caption{
      Ground state energies per site obtained for several $N$ with the Bethe Ansatz (BA) in the thermodynamic limit and on a finite size chain\cite{Martins_SUN}, and at $T=0.01$ with the quantum Monte-Carlo algortihm (see text) on a $L=60$ chain and $n=10^7$ Monte-Carlo steps.
      \label{tab:energies_SUN}
    }
  \end{table}

  \begin{figure*}
    \begin{center}
      \subfigure[\label{fig:correlations_chain_N2}\,$N=2$]{\includegraphics[width=0.24\textwidth,trim=3cm 1cm 2cm 2cm, clip]{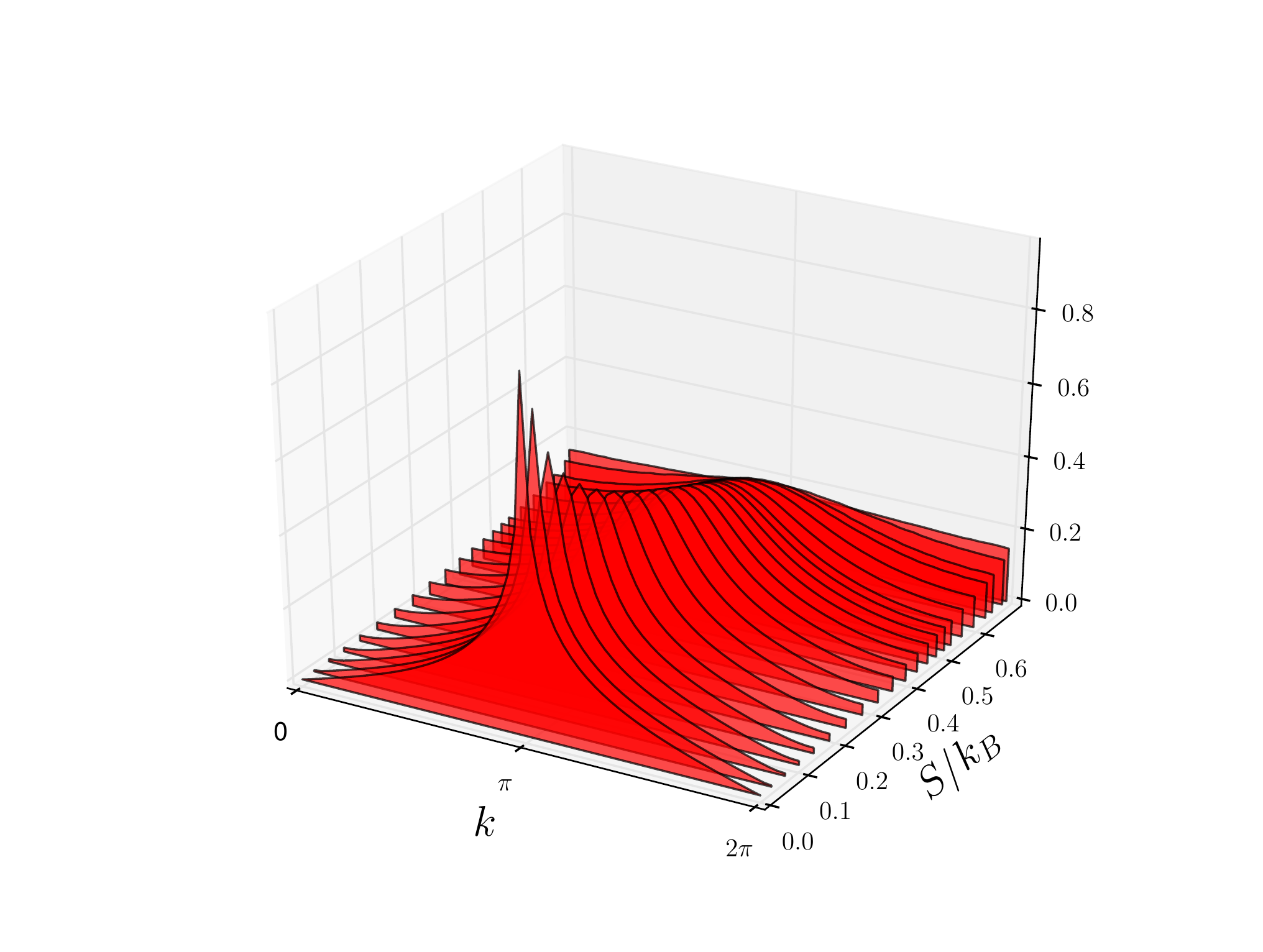}}
      \subfigure[\label{fig:correlations_chain_N3}\,$N=3$]{\includegraphics[width=0.24\textwidth,trim=3cm 1cm 2cm 2cm, clip]{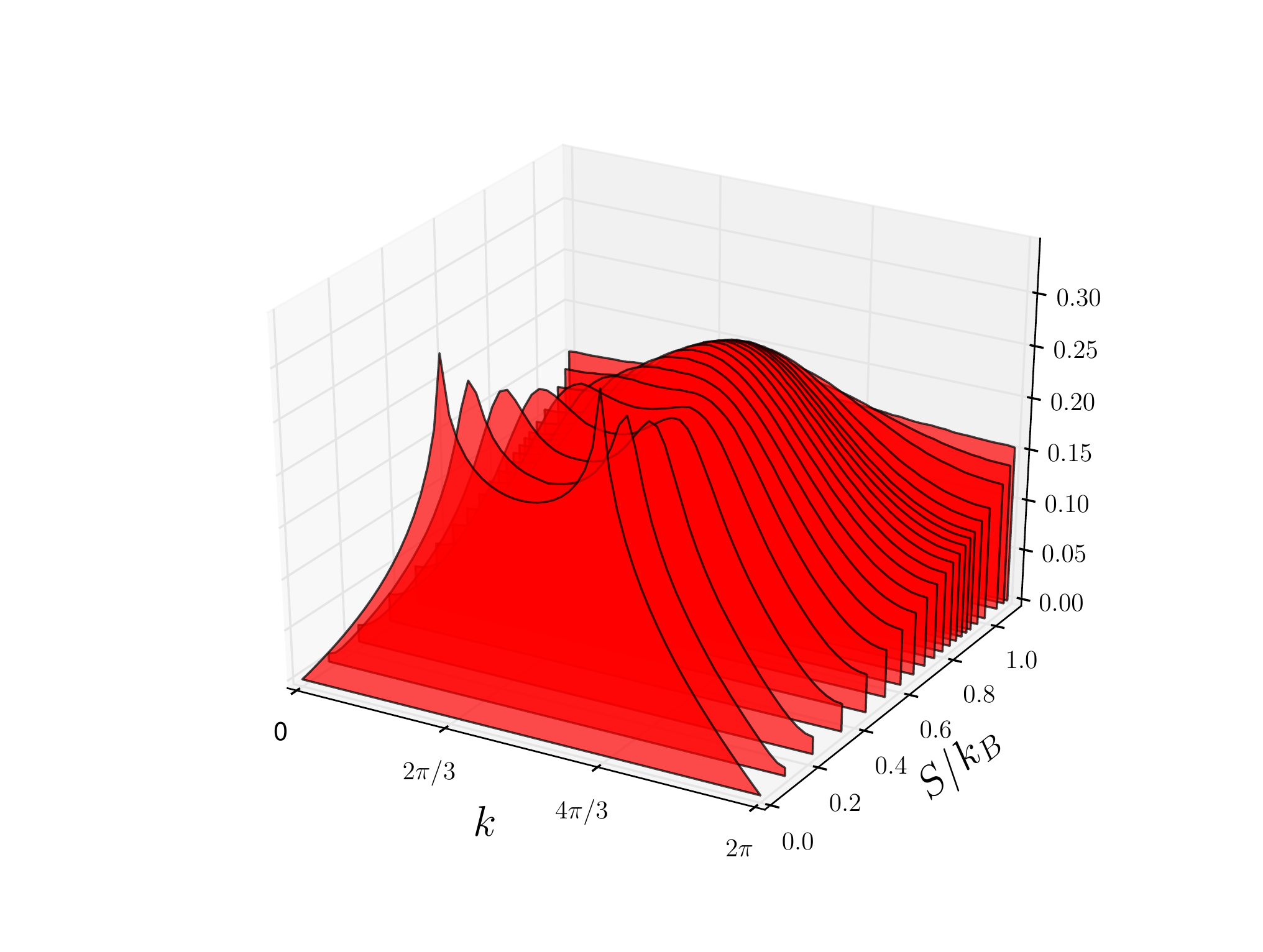}}
      \subfigure[\label{fig:correlations_chain_N4}\,$N=4$]{\includegraphics[width=0.24\textwidth,trim=3cm 1cm 2cm 2cm, clip]{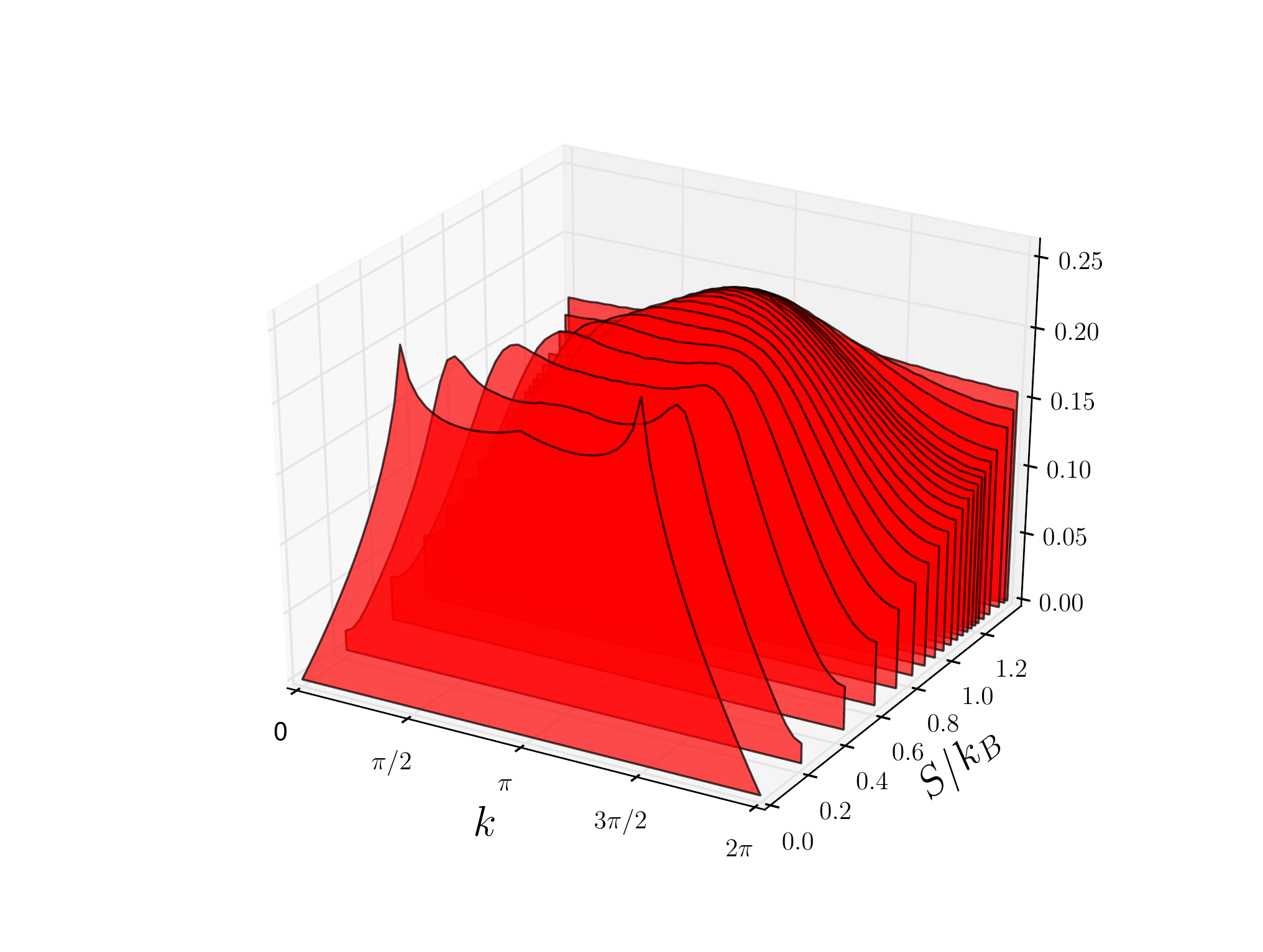}}
      \subfigure[\label{fig:correlations_chain_N5}\,$N=5$]{\includegraphics[width=0.24\textwidth,trim=3cm 1cm 2cm 2cm, clip]{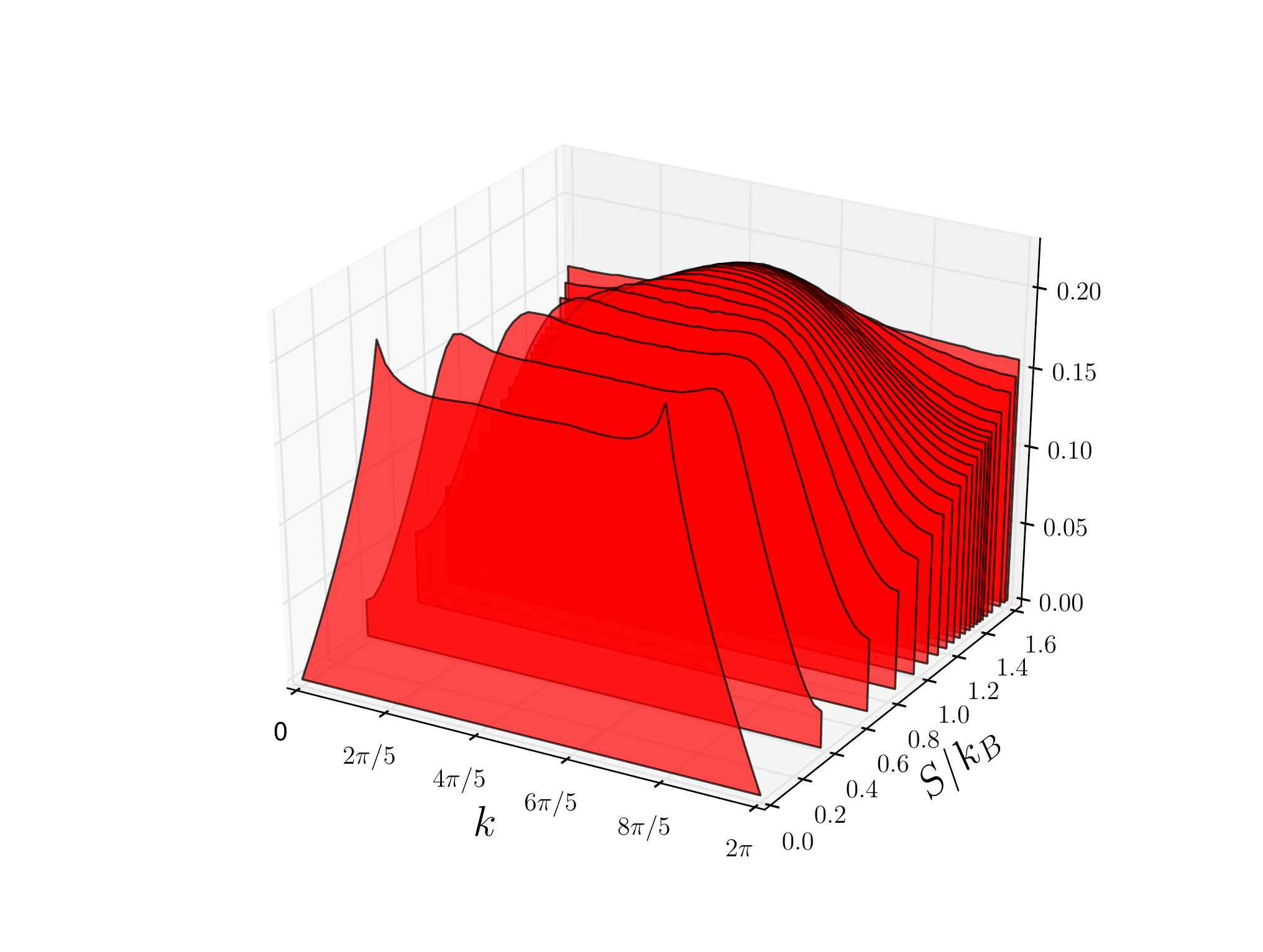}}
    \end{center}
    \caption{Evolution of the structure factor $\tilde C(k)$ as a function of the entropy per site $S$ for different $N$ on a $L=60$ chain.
    \label{fig:correlations_chain}}
  \end{figure*}

 \begin{figure}
  \begin{center}
    \includegraphics[trim=0 10 0 44,clip,width=.5\textwidth]{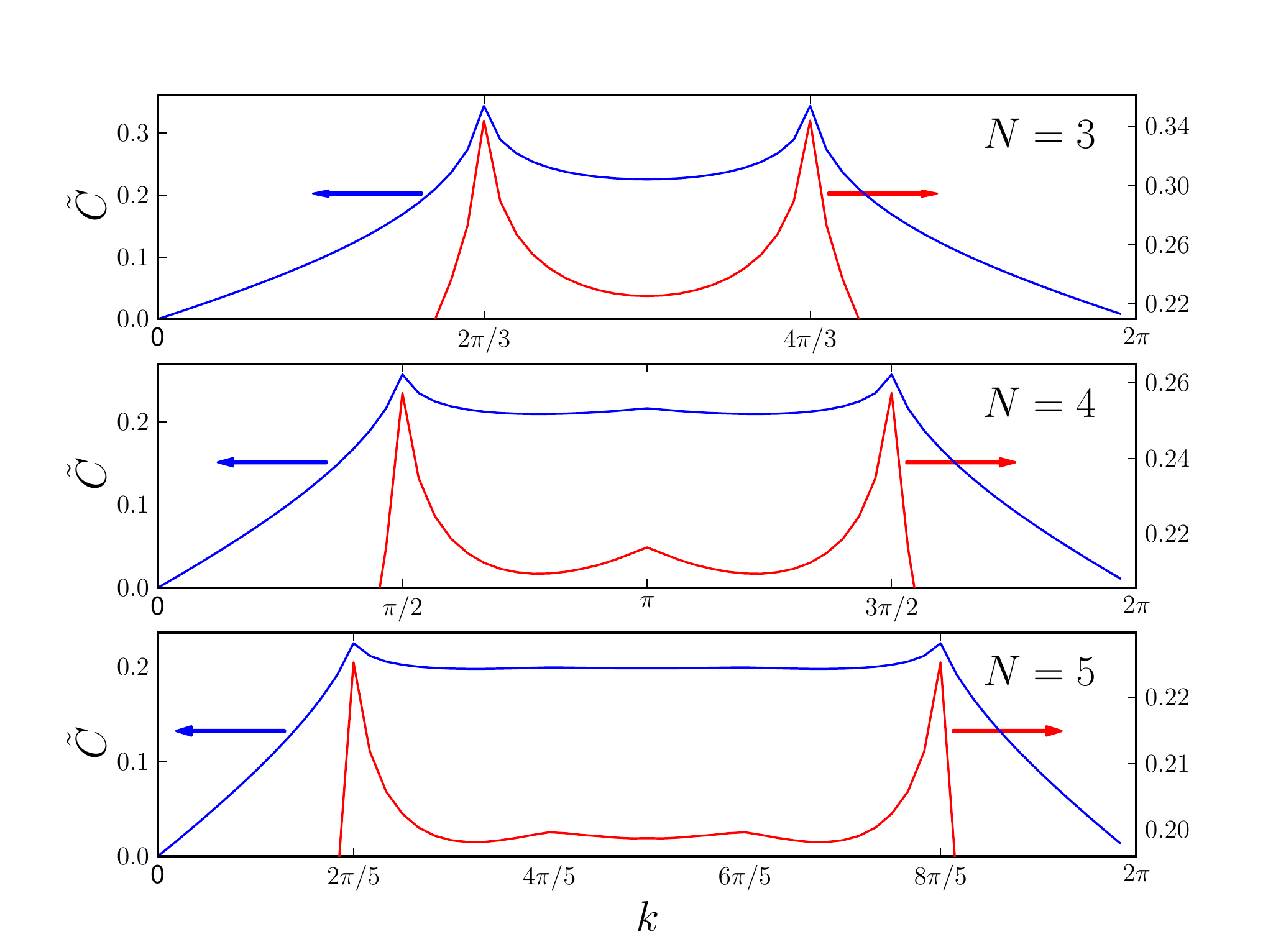}
  \end{center}
  \caption{Structures factor $\tilde C(k)$ at low temperature ($k_BT=0.01$) for different $N$ on a $L=60$ chain, with $n=10^7$ Monte-Carlo steps. Small peaks are clearly visible at $k=\pi$ for $N=4$ and $k=4\pi/5$ and $6\pi/5$ for
  $N=5$. The data for $N=4$ are in perfect agreement with those of Ref.\onlinecite{frischmuth_1999}.
  \label{fig:correlations_T001}}
  \end{figure}

  We now look at the diagonal correlations $\tilde C(k)$.
  They have been calculated for different temperatures, but, in view of the implications for
   ultracold fermionic gas, we represent them as a function of the entropy per site $S$.
  Since the system is 1D, there is no long range order, hence no Bragg peaks.
  Nevertheless, short-range correlation develop at low entropy. They translate into
  finite height peaks in $\tilde C(k)$ at finite temperature, and singularities at zero temperature.
  The number and the position of these peaks depend
  on the number of colors $N$. From the Bethe ansatz solution, singularities are expected to occur
  at $k=2p\pi/N$ with $p=1,...,N-1$. The results of Fig.~\ref{fig:correlations_chain} agree with
  this prediction: there is a single peak at $\pi$ for $SU(2)$, while $N-1$ peaks are indeed present
  for $SU(N)$ at sufficiently small entropy. Note however that all peaks do not have the same
amplitude for $N\geq4$.
  For $N=4$ and $5$, two types of peaks not related by the symmetry $k\rightarrow
  2\pi-k$ are present. The peaks
  at $2\pi/N$ and $2(N-1)\pi/N$ are much more prominent, and they
  start to be visible at much larger entropy.

At the maximal entropy, the structure factor $\tilde C(k)$ is flat (see Fig.~\ref{fig:correlations_chain}). At large but finite entropy, it presents a broad maximum at $k=\pi$ for all $N$.
This reflects the simple fact that colors tend to be different on neighboring sites. More specific correlations appear upon lowering
the entropy. For $SU(2)$, the peak at $k=\pi$ just gets more pronounced. To observe the development of the
singularity typical of the $SU(2)$ ground state algebraic correlations will however require to reach rather
low entropy. This should be contrasted with the $N>2$ cases, where a qualitative change in the structure
factor occurs upon reducing the entropy: the broad peak at $k=\pi$ is replaced by peaks at
$2\pi/N$ and $2(N-1)\pi/N$. One can in principle read off the corresponding
  entropy from Fig.~\ref{fig:correlations_chain}. To come up with
  a quantitative estimate, we note that, upon reducing the entropy, the curvature of the structure
  factor at $k=\pi$ changes sign from positive at high temperature to negative when the peaks at
$2\pi/N$ and $2(N-1)\pi/N$ appear
. This occurs at
$S_c/k_B=0.58$, $0.87$ and $1.08$ for $N=3$, $4$ and $5$ respectively.
This characteristic entropy $S_c$ increases more or less linearly with $N$ as $S_c \simeq 0.2 Nk_B$,
and for $N=4$ and $5$, it lies
in the experimentally accessible range. This is mostly a consequence of the temperature dependence of
the entropy, which grows much faster with $N$ at low temperature. The characteristic temperature
at which deviations from the broad peak at $k=\pi$ occur depends only weakly on $N$.
Finally, secondary peaks appear at lower temperature (see Fig.~\ref{fig:correlations_T001})

{\it Conclusions.---} We have shown that the entropy at which the periodicity characteristic of the zero
temperature algebraic order of $SU(N)$ chains is revealed increases significantly with $N$. For $N=4$, this
entropy is already larger than the entropy per particle recently achieved in the $N=2$ case in the center of
the Mott insulating cloud (0.77 $k_B$)\cite{jordens_2010}.
Whether a similar entropy can be achieved for $N>3$ remains to be seen. As shown by
Hazzard et al\cite{hazzard_2012}, if the initial
temperature is fixed, the initial entropy in a 3D trap increases with $N$ as $N^{1/3}$, implying that
one might have to go to values of $N$ larger than 4 to reach a final entropy low enough to observe
characteristic correlations. However, evaporative cooling might allow to reach initial entropies that
are less dependent on $N$. In a recent experiment on $^{173}$Yb, the initial entropy reported by Sugawa et al\cite{sugawa_2011} for this $N=6$ case is not much higher than in $N=2$ experiments\cite{jordens_2010}.
It is our hope that the present results will encourage the experimental
investigation of the $1/N$-filled Mott phase of $N$-color ultracold fermionic atoms.

We thank Daniel Greif for useful discussions. LM acknowledges the hospitality of EPFL, where most of this project has been
performed. This work has been supported by the Swiss National Fund and by MaNEP.

\bibliographystyle{apsrev4-1}
\bibliography{SUN_chain}

\end{document}